\documentstyle[aps,epsf,rotate,multicol]{revtex}

\begin{document}

\input psfig.tex
\title{Renormalization group of probabilistic   
cellular automata with one absorbing state}
\author{M.~J. de Oliveira and J.~E. Satulovsky}
\address{Universidade de S\~ao Paulo, \\ Instituto de F\'{\i}sica, 
Caixa Postal 66318, \\ 05315-970 S\~ao Paulo, SP, Brazil}
\date{\today}
\maketitle

\begin{abstract}
We apply a recently proposed dynamically driven renormalization group scheme
to probabilistic cellular automata having one absorbing state. We have found
just one unstable fixed point with one relevant direction. In the limit of
small transition probability one of the cellular automata reduces to the
contact process revealing that the cellular automata are in the same
universality class as that process, as expected. Better numerical results
are obtained as the approximations for the stationary distribution are
improved.
\end{abstract}

\begin{multicols}{2}
%%\narrowtext
%%%%%%%%%%%%%%%%%%%%%%%%%%%%%%%%%% INTRODUCTION %%%%%%%%%%%%%%%%%%%%%%%%
\section{Introduction}

Recently, a general dynamical renormalization group (RG) scheme \cite
{VesZapLor} has been proposed in order to treat nonequilibrium critical
phenomena. The method, called Dynamically Driven Renormalization Group
(DDRG), has been applied to self-organized critical phenomena, specifically
to critical height sandpile models \cite{VesZapPie} and forest fire models 
\cite{LorVesZap}. Its phenomenological approach takes into account the
nature of self-organized systems through an attractive fixed point. The
scheme also provides numerical values for the critical exponents which are
close to the ones obtained using computer simulations.

The scheme consists in coupling a real space RG scheme to a stationary
condition that drives the RG group equations through the parameter space.
The stationary conditions, involving the stationary distribution, must
always be approximated. While the authors used a stationary probability
distribution that neglects any correlation among different sites, they have
mentioned \cite{VesZapLor} the fact that using more refined approximations
should improve the values of the critical exponents.

While the DDRG scheme is quite general, self-organized critical systems are
special since they have a well defined time scale separation (dissipation
events being instantaneous with respect to the driving). This prevents
proliferation effects in the real space RG, making the calculation of
critical exponents easier.

In this work, we implemented the DDRG scheme to another class of systems
namely the class of driven diffusive systems with an absorbing state (the
vacuum state). Due to the presence of one absorbing state this class of
systems is in the same universality as the directed percolation model \cite
{Jans} \cite{Grass}. More precisely we have considered one-dimensional
probabilistic cellular automata with one absorbing state. We have treated
two types of models. One of them is a two-state probabilistic cellular
automaton (that can be interpreted as directed percolation in two dimensions)
that includes, in the limit of small transition probabilities, as particular
cases, continuous-time processes with one absorbing state such as the
contact model \cite{Lig} and others \cite{Ron}. The other one is a
four-state probabilistic cellular automaton that includes the model
introduced by Grassberger and de la Torre \cite{GrasTor} as a particular case.

The appropriate real space RG parameter space for these nonequilibrium
models is the space of the transition probabilities, instead of being the
space of coupling constants as is the case of equilibrium models, defined by
a Hamiltonian. In the case of the models studied here, with one absorbing
state, the RG should be appropriate to preserve the vacuum state along the RG
trajectory in this parameter space. By using a block renormalization to
treat properly the absorbing state we have figured the value of the critical
exponent corresponding to the divergence of the spatial correlation, $\nu
_{\perp }$, using three different approximations which consider correlations
among clusters up to $1$, $3$, and $5$ neighboring sites in the lattice,
respectively. Our best calculations give $\nu _{\perp }=1.04\pm 0.02$ which
is rather close to the one obtained from numerical simulations reported in 
\cite{GrasTor} namely $\nu _{\perp }=1.067\pm 0.005.$

%%%%%%%%%%%%%%%%%%%%%%%%%%%%%%%%%% TWO-STATE MODEL %%%%%%%%%%%%%%%%%%%%%
\section{Two-state model}

The first model is a one-dimensional cellular automaton with just two state
per site. Each site can be either vacant, $\sigma _i=0,$ or occupied by a
particle, $\sigma _i=1.$ At each time step the state of a certain site will
depend only on the previous states of that same site and its nearest
neighbors. We consider the most general transition probabilities that are
homogeneous and symmetric in space. The transition probability $W(\sigma
|\sigma ^{\prime })$ from state $\sigma ^{\prime }=(\sigma _1^{\prime
},\sigma _2^{\prime },\cdots,\sigma _N^{\prime })$ to state $\sigma =(\sigma
_1,\sigma _2,\cdots ,\sigma _N)$ will be given by the product 
\begin{equation}
W(\sigma |\sigma ^{\prime })=\prod_{i=1}^Nw(\sigma _i|\sigma _{i-1}^{\prime
},\sigma _i^{\prime },\sigma _{i+1}^{\prime }),  \label{1}
\end{equation}
where $N$ is the number of sites and $w(\sigma _i|\sigma _{i-1}^{\prime
},\sigma _i^{\prime },\sigma _{i+1}^{\prime })$ is the one-site transition
probability given by the rules 
\begin{equation}
\label{2}
\begin{array}{|c|c|c|c|c|c|c|c|c|} \hline
w & 000 & 001 & 100 & 101 & 010 & 011 & 110 & 111 \\ \hline
0 & 1 & 1\!\!-\!\!p_1 & 1\!\!-\!\!p_1 & 1\!\!-\!\!p_2 & 
p_3 & p_4 & p_4 &p_5 \\ \hline
1 & 0 & p_1 & p_1 & p_2 & 1\!\!-\!\!p_3 & 1\!\!-\!\!p_4 & 
1\!\!-\!\!p_4 & 1\!\!-\!\!p_5 \\ \hline
\end{array}
\end{equation}

The rule $w(0|000)=1$ implies that the vacuum state is indeed an absorbing
state. When $p_5=p_4=p_3$ we say that the annihilation of particles is
spontaneous. Suppose moreover that $p_2=2p_1$ and that the parameters $p_1$
and $p_3$ are very small. In this case the system remains most of the time
in its previous state. We expect, therefore, that the properties of the
present two-state model, in the limit $p_1\rightarrow 0$ and $p_3\rightarrow
0,$ with the ratio $p_1/p_3=\lambda $ fixed, be identical to the contact
process with a catalytic transition rate equal to $\lambda $ and a
transition rate for spontaneous annihilation equal to unity. If $p_2=p_1,$
and taking the same limit, the properties will be identical to a model
introduced by Dickman (Model A) \cite{Ron}.

The model is represented by a set of five parameters, $p_1,$ $p_2,$ $p_3,$ $%
p_4,$ and $p_5,$ which constitutes, as we shall see, the RG parameter space.
The RG scheme will be constructed in a way that the RG trajectory will be
confined to this space.

%%%%%%%%%%%%%%%%%%%%%%%%%%%%%%%%%% FOUR-STATE MODEL %%%%%%%%%%%%%%%%%%%%%
\section{Four-state model}

Grassberger and the la Torre model \cite{GrasTor} is defined as follows.
Each site of a one-dimensional lattice is either occupied by one particle or
it is void. At a certain time step the state of the system may be defined by
the vector $\sigma =(\sigma _1,\sigma _2,\sigma _3,\cdots ,\sigma _N)$ where 
$\sigma _i=0$ or $1$ according whether the site $i$ is vacant or occupied. In
each time step, all sites are updated in two stages.
\begin{itemize}
 \item[1)] In the first stage each particle is spontaneously annihilated with
probability $c$.
 \item[2)] In the second stage every surviving particle will generate, with
probability $p,$ a new (unique) particle, which will be placed in one of its
nearest neighboring sites, randomly chosen. In other words, for each site
with a particle, we chose a neighboring site with probability $p/2$. If the
site was originally void it becomes occupied, and if it was occupied it
remains as such. We have modified slightly the original model by introducing
the parameter $p.$ The original model of Grassberger and de la Torre is recast
when $p=1$.
\end{itemize}

Defined in this way, the transition probability $W(\sigma |\sigma ^{\prime
}) $ from a state $\sigma ^{\prime }$, to another state $\sigma $, can not
be written as a product of independent transition probabilities associated
to each site $w(\sigma _i|\sigma _{i-1}^{\prime },\sigma _i^{\prime },\sigma
_{i+1}^{\prime })$, as in ordinary cellular automaton. However, if we
enlarge the number of states in each site by introducing three types of
particle, then it is possible to map the model into a four-state cellular
automata. This mapping is outlined in Appendix A.

The four-state probabilistic cellular automaton equivalent to Grassberger 
and de la Torre model is defined as follows. Each site of a one-dimensional
lattice can be either empty ($E$), $\sigma _i=0,$ or occupied by a neutral
particle ($N$), $\sigma _i=1,$ or by a rightist particle ($R$), $\sigma
_i=2, $ or by a leftist particle ($L$), $\sigma _i=3.$ At each time step,
every site of the lattice is independently updated according to the rules.
\begin{itemize}
\item[1)] If the site is occupied by one particle of any type $N$, $R$, 
or $L$, then one out of four possible events will take place.
\begin{itemize}
\item[a)] The particle is annihilated, that is, the site becomes empty, with
probability $c$, or
\item[b)] becomes a particle of type $N$ with probability $a$, or
\item[c)] becomes a particle of type $R$ with probability $b/2$, or
\item[d)] becomes a particle of type $L$ with probability $b/2$.
\end{itemize}
Here $a=(1-c)(1-p)$ and $b=(1-c)p.$

\item[2)] In case the site is empty (state $E$) one has to look to its 
neighboring sites.
\begin{itemize}
\item[a)] If its left neighbor is type $R$ or its right neighbor is of type 
$L$, the site remains as $E$ with probability $c$, becomes $N$ with 
probability $a $, becomes $R$ with probability $b/2$, or becomes 
$L$ with probability $b/2 $.
\item[b)] If on the contrary, its left nearest neighbor is not a particle of type $R$ and is right nearest neighbor is not of type $L$, the site remains
vacant. This last rule implies that the state with all sites empty is an
absorbing state.
\end{itemize}
\end{itemize}

Transition probability $W(\sigma |\sigma ^{\prime })$ from state $\sigma
^{\prime }=(\sigma _1^{\prime },\sigma _2^{\prime },\cdots,\sigma _N^{\prime 
})$ to state $\sigma =(\sigma _1,\sigma _2,\cdots ,\sigma _N)$ can be written 
as the product 
\begin{equation}
W(\sigma |\sigma ^{\prime })=\prod_{i=1}^Nw(\sigma _i|\sigma _{i-1}^{\prime
},\sigma _i^{\prime },\sigma _{i+1}^{\prime }),  \label{3}
\end{equation}
where $N$ is the number of sites of the lattice. The one-site transition
probability $w(\sigma _i|\sigma _{i-1}^{\prime },\sigma _i^{\prime },\sigma
_{i+1}^{\prime })$ is written down in the Appendix A.

In order to apply the RG scheme we enlarge the space of parameters but
preserve the existence of the absorbing state. Demanding besides that the
rules should be homogeneous in space and invariant by exchanging the states 
$R$ and $L,$ the transition probability $w(\sigma _i|\sigma _{i-1}^{\prime
},\sigma _i^{\prime },\sigma _{i+1}^{\prime })$ will be defined in the most
general way by $59$ parameters. We call such a probabilistic cellular
automaton the four-state model.

%%%%%%%%%%%%%%%%%%%%%%%%%%%%%%%%%% RENORMALIZATION SCHEME %%%%%%%%%%%%%%%%%
\section{Renormalization Scheme}

Here we use a real space RG scheme \cite{MazNolVal,MazVal,Haake} 
which renormalizes the transition probability $W.$ The succession of
RG transformations corresponds to a trajectory in the space spanned by the
parameters that defines $W.$ The scheme we use is an implementation of 
the DDRG \cite{VesZapLor,VesZapPie,LorVesZap} and is accomplished by
transforming cells of $b$ sites into a cell of just one site. To treat the
vacuum state properly any cell with at least one particle renormalizes into
an occupied site. Only cells with no particles renormalizes into a vacant
site.

Let ${\cal R}(\tau |\sigma )$ be a condition probability of state $\tau $
given the state $\sigma $ with the following properties 
\begin{equation}
{\cal R}(\tau |\sigma )\geq 0,\qquad \qquad \sum_\tau {\cal R}(\tau |\sigma
)=1.  \label{4}
\end{equation}
The vector $\sigma =(\sigma _1,\sigma _2,\cdots ,\sigma _N)$ represents the
state of a system with $N$ degrees of freedom and the vector $\tau =(\tau
_1,\tau _2,\cdots ,\tau _{N^{\prime }})$ represents the state of the
renormalized system with $N^{\prime }=N/b$ degrees of freedom, where $b$ is
the size of the renormalization block.

Let $P_n(\sigma ,\sigma ^{\prime })$ be the probability of occurrence of
state $\sigma ^{\prime }$ at a given time and state $\sigma $ at $n$ time
steps later, that is 
\begin{equation}
P_n(\sigma ,\sigma ^{\prime })=W^n(\sigma |\sigma ^{\prime })
P(\sigma ^{\prime })
\label{5}
\end{equation}
where $P(\sigma )$ is the stationary probability distribution which
satisfies the equation 
\begin{equation}
P(\sigma )=\sum_{\sigma ^{\prime }}W^n(\sigma |\sigma ^{\prime })
P(\sigma ^{\prime })
\label{6}
\end{equation}
for any value of $n$, where $W^n(\sigma |\sigma ^{\prime })$ is the
transition probability from state $\sigma ^{\prime }$ to state $\sigma $ in $%
n$ time steps. Similarly, for the renormalized system, let $\widetilde{P}%
(\tau ,\tau ^{\prime })$ be the probability of occurrence of state $\tau
^{\prime }$ at a given time and state $\tau $ at one time step later. The RG
transformation are obtained by demanding that \cite{MazNolVal} 
\begin{equation}
\widetilde{P}(\tau ,\tau ^{\prime })=\sum_\sigma \sum_{\sigma ^{\prime }}%
{\cal R}(\tau |\sigma ){\cal R}(\tau ^{\prime }|\sigma ^{\prime })P_n(\sigma
,\sigma ^{\prime })  \label{7}
\end{equation}
from which follows 
\begin{equation}
\widetilde{P}(\tau ^{\prime })=\sum_{\sigma ^{\prime }}{\cal R}(\tau
^{\prime }|\sigma ^{\prime })P(\sigma ^{\prime })  \label{8}
\end{equation}
since 
\begin{equation}
\widetilde{P}(\tau ^{\prime })=\sum_\tau \widetilde{P}(\tau ,\tau ^{\prime })
\label{9}
\end{equation}

To get the desired renormalized transition probability $\widetilde{W}(\tau
|\tau ^{\prime })$ we use 
\begin{equation}
\widetilde{W}(\tau |\tau ^{\prime })=\frac{\widetilde{P}(\tau ,\tau ^{\prime
})}{\widetilde{P}(\tau ^{\prime })},  \label{10}
\end{equation}
and equations (\ref{5}), (\ref{7}), and (\ref{8}). 
We obtain \cite{VesZapLor} 
\begin{equation}
\widetilde{W}(\tau |\tau ^{\prime })=\frac{\sum_\sigma \sum_{\sigma
^{\prime }}{\cal R}(\tau |\sigma ){\cal R}(\tau ^{\prime }|\sigma ^{\prime
})W^n(\sigma |\sigma ^{\prime })P(\sigma ^{\prime })}
{\sum_{\sigma ^{\prime }}{\cal R}
(\tau ^{\prime }|\sigma ^{\prime })P(\sigma ^{\prime })}  \label{11}
\end{equation}
This equation, however, is not properly a transformation between the 
transition probabilities $W$ and $\widetilde{W}$, since the yet unknown 
stationary probability $P(\sigma )$ appears in the right hand side of equation
(\ref{11}). However, if we use the balance equation (\ref{6}) for $P(\sigma )$ 
then a closure condition for the renormalization group is obtained.
The closure condition plays the role of the driving condition that 
forces the system to be in the stationary state at each step of the 
transformation. In this sense the present DDRG may be thought of as 
a renormalization of the stationary state.

At each state of the transformation the transition probability $W$ 
always describes an irreversible process so that the corresponding  
stationary solution $P(\sigma )$ obtained from the closure condition 
(\ref{6}) will not be related, a priori, to a Hamiltonian, that would 
be the case if the process obeyed detailed balance. In this way the 
present RG scheme is distinct from the ordinary real space RG used in
equilibrium systems in which the parameters of the Hamiltonian are 
renormalized.

The closure relation (\ref{6}), however, cannot actually be solved so that
approximations should be used. Here we have used three different 
approximations which consider correlations among clusters up to 
$1$, $3$, and $5$ neighboring sites. Equation (\ref{11}) together with 
a given approximation provides then a well defined RG transformation 
$W \rightarrow \widetilde{W}$.

Assuming that the renormalized transition probability can also be written as
a product of independent transition probabilities, that is, 
\begin{equation}
\widetilde{W}(\tau |\tau ^{\prime })=\prod_{k=1}^{N^{\prime }}\widetilde{w}%
(\tau _k|\tau _{k-1}^{\prime },\tau _k^{\prime },\tau _{k+1}^{\prime }),
\label{16}
\end{equation}
one can write the RG equations that relates the old one-site transition
probability $w(\sigma _i|\sigma _{i-1}^{\prime },\sigma _i^{\prime },\sigma
_{i+1}^{\prime })$ to the new one $\widetilde{w}(\tau _k|\tau _{k-1}^{\prime
},\tau _k^{\prime },\tau _{k+1}^{\prime }).$

We have used only renormalization cells with size $b=2$ and chosen ${\cal R}$
in the form 
\begin{equation}
{\cal R}(\tau |\sigma )=\prod_{k=1}^{N/2}R(\tau _k|\sigma _{2k-1},\sigma
_{2k}),  \label{21}
\end{equation}
with 
\begin{equation}
R(\tau _k|\sigma _{2k-1},\sigma _{2k})\geq 0,
\mbox{and } \qquad \sum_{\tau
_k}R(\tau _k|\sigma _{2k-1},\sigma _{2k})=1  \label{22}
\end{equation}
To preserve the absorbing nature of the vacuum state we have chosen $R$ with
the properties 
\begin{equation}
R(0|0,0)=1  \label{23}
\end{equation}
and 
\begin{equation}
R(0|\sigma _{2k-1},\sigma _{2k})=0  \label{24}
\end{equation}
whenever $\sigma _{2k-1}\neq 0,$ or $\sigma _{2k}\neq 0.$

For the four-site model we used the following values 
$$
\begin{array}{cccc}
R(0|00)=1 & R(1|01)=1 & R(1|10)=1 & R(1|11)=1 \\ 
R(2|02)=1 & R(1|03)=1 & R(1|20)=1 & R(3|30)=1 \\ 
R(2|12)=1 & R(1|13)=1 & R(1|21)=1 & R(3|31)=1 \\ 
R(2|32)=1/2 & R(3|32)=1/2 & R(2|22)=1 & R(1|23)=1
\end{array}
$$
The other matrix elements of $R$ are zero. The first matrix element has been
chosen in order to preserve the absorbing nature of the vacuum state and the
rest of the elements where assigned bearing in mind the physical picture
that an $R$ particle will give rise to a new particle to the right in the
next time step, while an $L$ particle will generate a new particle to the
left ($N$ particles will not generate new particles).

%%%%%%%%%%%%%%%%%%%%%%%%%%%%%%%%%% RENORMALIZATION ALGORITHM %%%%%%%%%%%%%%
\section{Renormalization algorithm}

\begin{figure}
\narrowtext
\centerline{\rotate[r]{\epsfysize=3.3in \epsffile{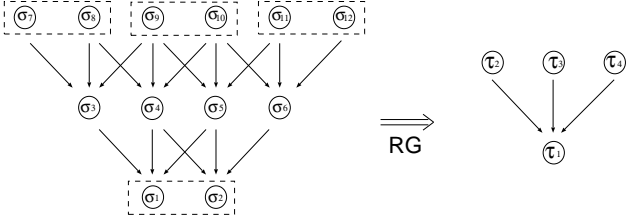}}}
\caption{Diagram showing the blocking scheme procedure. Numbers
correspond to the indexes used in equation~(\ref{31}).}
\label{fig1}
\end{figure}

The temporal coarse graining will be done using two time steps that is $n=2.$
Using the equations of the previous section we can write down the equation
that relates $w$ to $\widetilde{w}$ for the case $n=2$ (see figure~\ref{fig1}).

It is given by 
\begin{eqnarray}
\widetilde{w}(\tau _1|&&\tau _2,\tau _3,\tau _4)=\!\!\!\!\!\!\!
\sum_{\sigma _1 \sigma _2 \sigma _7 \cdots \sigma _{12}}\!\!\!\!\!\!\!
R(\tau _1|\sigma _1\sigma _2)\times \nonumber \\ 
&&
T(\sigma _1,\sigma _2|\sigma _7,\cdots ,\sigma _{12})
\rho (\sigma_7,\cdots ,\sigma _{12}|\tau _2,\tau _3,\tau _4),  
\label{31}
\end{eqnarray}
where 
\begin{eqnarray}
T(&&\sigma _1\sigma _2|\sigma _7,\sigma _8,\sigma _9,\sigma _{10},\sigma
_{11},\sigma _{12})=\nonumber \\
&&\!\!\!\!\!\!\sum_{\sigma _3 \sigma _4 \sigma _5 \sigma _6}\!\!\!\!\!\!
w(\sigma _1|\sigma _3,\sigma _4,\sigma _5)
w(\sigma _2|\sigma _4,\sigma _5,\sigma _6) \times \nonumber \\
&&~~~~~~w(\sigma _3|\sigma _7,\sigma _8,\sigma _9)
w(\sigma _4|\sigma _8,\sigma _9,\sigma _{10})\times \nonumber \\  
&&~~~~~~w(\sigma _5|\sigma _9,\sigma _{10},\sigma _{11})
w(\sigma _6|\sigma _{10},\sigma _{11},\sigma _{12})^{\mbox{\ }},  \label{32}
\end{eqnarray}
and 
\begin{eqnarray}
\rho (\sigma _7&,&\sigma _8,\sigma _9,\sigma _{10},\sigma _{11},
\sigma _{12}|\tau _2,\tau _3,\tau _4)=
{[\widetilde{P}(\tau _2,\tau _3,\tau _4)]}^{-1}\times \nonumber \\ 
&& 
R(\tau _2|\sigma _7,\sigma _8) R(\tau _3|\sigma _9,\sigma _{10})
R(\tau _4|\sigma _{11},\sigma _{12})\times \nonumber \\ 
&&
P(\sigma _7,\sigma _8,\sigma _9,\sigma _{10},\sigma _{11},\sigma _{12})
^{\mbox{\ }},
\label{33}
\end{eqnarray}
where 
\begin{eqnarray}
\widetilde{P}(\tau _2,&\tau _3&,\tau _4)= \!\!\!\!\!
\sum_{\sigma _7 \cdots \sigma _{12}}\!\!\!\!\! 
R(\tau _2|\sigma _7,\sigma _8) R(\tau _3|\sigma _9,\sigma _{10})
\times \nonumber \\ 
&&R(\tau _4|\sigma _{11},\sigma _{12}) 
P(\sigma _7,\sigma _8,\sigma _9,\sigma _{10},\sigma _{11},\sigma _{12}).  
\label{34}
\end{eqnarray}
Here the subscripts refer to the site numbers appearing in figure~\ref{fig1}.

In order to solve the system of equations (\ref{31}-\ref{34}) one must
resort on approximate methods to estimate the stationary weights $P(\sigma
_7,\cdots,\sigma _{12}).$ The simplest approximation, sometimes known as simple
mean field approximation, consists in neglecting correlations among
different sites

\begin{equation}
P(\sigma _7,\cdots ,\sigma _{12})=\prod_{i=7}^{12}P(\sigma _i),  \label{41}
\end{equation}
where $P(\sigma _i)$ is the solution of 
\begin{equation}
P(\sigma _1)=\sum_{\sigma _2\sigma _3\sigma _4}w(\sigma _1|\sigma _2\sigma
_3\sigma _4)P(\sigma _2)P(\sigma _3)P(\sigma _4).  \label{42}
\end{equation}

However, correlations are actually taken into account in the geometrical
aspects of the blocking procedure. In this way, one obtains non classical
critical exponents. Given a blocking prescription, the value of the critical
exponents should improve as correlations are taken into account into the
stationary probability distribution. In order to verify how important the
changes will be, we have used three different approximations for the
stationary distribution. The first one being (\ref{41}-\ref{42}), while the
other two are mean field approximations which consider correlations up to
clusters of three sites and five sites respectively \cite{mfield}.

Due to the number of terms involved in equations (\ref{31}-\ref{34}) it is
not possible to determine analytically the fixed point of the
transformation. So, we have performed it numerically, using initial values
for the transition probabilities corresponding to the model of interest. In
the case of the two-state model we start with $w(\sigma _i|\sigma
_{i-1}^{\prime },\sigma _i^{\prime },\sigma _{i+1}^{\prime })$ given by
(\ref{2}) with $p_5=p_4=p_3$ and $p_2=2p_1$. In the case of the
four-state model we start with $w(\sigma _i|\sigma _{i-1}^{\prime },\sigma
_i^{\prime },\sigma _{i+1}^{\prime })$ corresponding to the Grassberger 
and de la Torre model, with $p=1$.

In each iteration of the RG, given the set of parameters $\widetilde{w}$,
one has to find the stationary solution for $P(\sigma _7,\cdots,\sigma _{12})$.
This has been done by iterating the time evolution equation for the model
(using one of the three approximations) until reaching convergence. For the
one-site approximation, given by (\ref{42}), $10^4$ iterations were enough.
As approximations are refined equations become highly nonlinear, and for the
five-site approximation least $10^5$ iterations are needed.

%%%%%%%%%%%%%%%%%%%%%%%%%%%%%%%%%% RESULTS %%%%%%%%%%%%%%%%%%%%%
\section{Results}

For the four-state model, the RG equations behaved in the following way. For
small values of $c$, the set of transition matrix elements flows towards an
attractive fixed point characterized by $c=0$, and a lattice full of
particles. On the contrary, for values of $c$ high enough, the RG equations
are driven to a different attractive fixed point, this time characterized by 
$c=1$ and a lattice without particles. In this case we used only the
one-site approximation and found a critical value of $c$ given by 
$c_{cr}=0.3568.$ Starting around this values the representative point of the
parameter set spends a long time near an unstable point before it leaves
towards any of the two attractive fixed points. Figure 2 shows a projection
of two trajectories in the parameter space in terms of two of these
parameters: $w(0|010)$ and $1-w(0|101)$.

\begin{figure}
\narrowtext
\centerline{\rotate[r]{\epsfysize=3.7in \epsffile{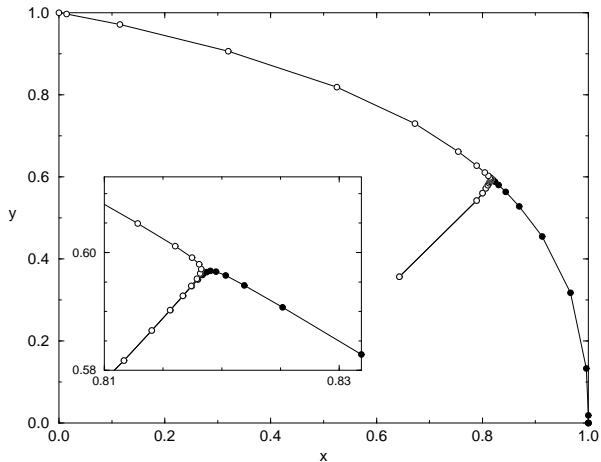}}}
\caption{Projection of two trajectories in the parameter space 
of the four-state model in terms of 
two of these parameters: $x=1-\widetilde{w}(0|101)$ and 
$y=\widetilde{w}(0|010)$. Each trajectory reaches a different 
fixed point. The inset shows an enlargement around the 
nontrivial fixed point.}
\label{fig2}
\end{figure}

In this way, it is found only one relevant parameter. Since we are dealing
only with stationary properties of the model it is reasonable to assume that
this parameter is associated to the divergence of the spatial correlation
length and not to the temporal correlation length. So, figuring the
eigenvalue $\Lambda$ associated to that parameter we get $\nu _{\perp }=
\ln 2/\ln \Lambda $. The value measured numerically is $\nu _{\perp }=0.93
\pm 0.005$.

From computer simulations results reported in \cite{GrasTor} one can obtain 
$\nu _{\perp }=1.067\pm 0.005$, and the critical value of $c$ is $c_{cr}=
0.279$. The discrepancy between the two results are mainly due to the 
poorness of the one-site approximation. By increasing the order of 
approximation the results gets better as we shall see in the case of the
two-site model.

We have corroborated that, as one would expect, the value of $\nu _{\perp }$
for the simpler two-state model and one-site approximation is identical as
the one previously found. Since for refined approximations of three and five
sites, numerical computations become too imposing, we have used these two
better approximations only for the two-state model. The value found using
the three-site approximation was $\nu _{\perp }=0.98\pm 0.01$ whereas the
value found in the five-site approximation was $\nu _{\perp }=1.04\pm 0.02$,
which is rather close to the one calculated from numerical simulations.
Bellow we show the coordinates of the unstable fixed point for the three
approximations, as well as the corresponding value of $\Lambda$
\[
\begin{array}{|c|c|c|c|c|c|c|} \hline
\text{appr.}\!& p_1\!& p_2 \!& p_3  \! & p_4  \! & p_5 \!  & \Lambda \\ \hline
\text{1} & 0.22794 & 0.40381 & 0.26780 & 0.22116 & 0.18245 & 2.105 \\ \hline
\text{3} & 0.12107 & 0.22750 & 0.14761 & 0.07870 & 0.03140 & 2.025 \\ \hline
\text{5} & 0.07107 & 0.16233 & 0.07330 & 0.04783 & 0.01933 & 1.950 \\ \hline
\end{array}
\]
Another advantage in using better approximations is the improvements one
obtains in the critical values of non-universal parameters. We have consider
the following initial conditions for the two-state model: $p_5=p_4=p_3$ and 
$p_2=2p_1.$ By varying $p_1$ and $p_3$ we have obtained the critical values 
$\lambda _c$ of the ratio $\lambda =p_1/p_3$ as shown in the table 
\begin{equation}
\begin{array}{|c|c|c|c|c|} \hline
p_1 & 1 & 0.1 & 0.01 & 0.001 \\ \hline
\lambda _c\quad \text{(1-site app.)} & 2.8027 & 2.0833 & 2.0088 & 2.0008 \\ \hline
\lambda _c\quad \text{(3-site app.)} & 3.2365 & 3.1205 & 3.1703 & 3.1757 \\ \hline
\lambda _c\quad \text{(5-site app.)} & 3.2555 & 3.1645 & 3.2206 & 3.2300 \\ \hline
\end{array}
\end{equation}
which, in the limit $p_1\rightarrow 0,$ should be compared with the critical
value of the contact process $\lambda _c=3.299$ \cite{BrowFurMosh}. The fact
that the column associated to $p_1=1$ also seems to converge to this value
is a mere coincidence.

Reasoning along this line, one may wonder why not to use the stationary
distribution directly from a computer simulation of the model. Making large
statistics over configurations in the stationary regime, one should be able
to accurately estimate the probabilities of the clusters appearing on the left
hand side of equations (\ref{33}) and (\ref{34}). While the idea is in
principle right, one can not overcome in practice the huge amount of time
needed to obtain values that are accurate enough. Slight fluctuations in the
estimated values will make trajectories randomly shift their destination
towards one of the two attractive fixed points, depending on the random seed
used in the simulation. This effect takes place even before any trajectory
is able to reach the linearized domain of the transformation around the
unstable point. A way to decrease fluctuations is increasing the size of the
lattice in which one performs the simulation. But the precision one gains
does not grow faster than $\sim 1/\sqrt{N}$, where $N$ is the size of the
lattice.

%%%%%%%%%%%%%%%%%%%%%%%%%%%%%%%%%% CONCLUSION %%%%%%%%%%%%%%%%%%%%%
\section{Conclusion}

We have applied a real space renormalization group scheme to a class of
driven diffusive probabilistic cellular automata having one absorbing state.
Two models having been considered. One of them is a two-state model that
reduces to the contact process in the limit of small transition
probabilities. The two-state model can also be interpreted as a generic
directed percolation in two dimensions. The other is a four-state models that
includes the model introduced by Grassberger and de la Torre in a study
related to the contact process. We have found, in the RG space of
parameters, just one non-trivial unstable fixed point with one relevant
direction. The existence of just this unique fixed point reveals that the
probabilistic cellular automata with one absorbing state belongs to the same
universality class as the directed percolation and the contact process, as
expected.

The implementation of the RG scheme nedeed the calculation of the stationary
probability distribution which were performed in several levels of
approximations. Increasing the number of cluster size used in the
approximation improved results were obtained not only for the critical
exponent $\nu _{\bot }$ as well as for the non-universal critical
quantities. 

%%%%%%%%%%%%%%%%%%%%%%%%%%%%%%%%%% APPENDIX A %%%%%%%%%%%%%%%%%%%%%
\section{Appendix A}

The Grassberger and de la Torre model is a stochastic process in which at each
time step particles are being created and annihilated. We imagine the
process as a sequence os states $A,$ $A^{\prime },$ $A^{\prime \prime },$ $%
A^{\prime \prime \prime }, \cdots $ each one being given by a vector $\eta
=(\eta _1,\eta _2,\cdots,\eta _N)$ where $\eta _i=0$ or $1$ according whether
site $i$ is vacant or occupied by a particle. We may think of each
transition, say $A\rightarrow A^{\prime },$ as composed of three stages with
two intermediate states $B$ and $C,$ to be defined shortly, so that the
whole stochastic process corresponds to a sequence $A,$ $B,$ $C,$ $A^{\prime
},$ $B^{\prime },$ $C^{\prime },$ $A^{\prime \prime },$ $B^{\prime \prime },$
$C^{\prime \prime },$ $A^{\prime \prime \prime },\cdots $ We will then write
the transition probability $W(A^{\prime }|A)$ from state $A$ to $A^{\prime }$
as given by 
\begin{equation}
W(A^{\prime }|A)=\sum_B\sum_CW_3(A^{\prime }|C)W_2(C|B)W_1(B|A).
\label{a1}
\end{equation}
where $W_3(A^{\prime }|C),$ $W_2(C|B),$ and $W_1(B|A)$ are the intermediate
transition probabilities related to the three stages.

{\em First stage}. In the first stage ($A\rightarrow B$) of the Grassberger
and de la Torre model, each particle is annihilated with probability $c$, so
that the probability $W_1(B|A)$ of the transition from $A$ to $B$ is given
by 
\begin{equation}
W_1(B|A)=W_1(\eta ^{\prime \prime }|\eta )=\prod_{i=1}^Nw_1(\eta _i^{\prime
\prime }|\eta _i),  \label{a2}
\end{equation}
where 
\begin{equation}
\begin{array}{ll}
w_1(0|1)=c, & \quad w_1(1|1)=1-c, \\
w_1(1|0)=0, & \quad w_1(0|0)=1.
\label{a3}
\end{array}
\end{equation}
{\em Second stage}. In this stage ($B\rightarrow C$) every particle decides
whether it will generate new particles either to the left or to the right in
the next step. Each occupied site will be labeled according to its decision.
A particle that decides not to generate particles will be labeled by the
number $1.$ A particle that decides to generate another one to the right,
will be labeled by the number $2$ and a particle that decides to generate a
new particle to the left, will be labeled by the number $3$. A state of type 
$C$ is then defined by the vector $\sigma =(\sigma _1,\sigma _2,\cdots ,\sigma
_N) $ where $\sigma _i=0,1,2,3,$ so that the transition probability $W_2(C|B)
$ of the transition from $B$ to $C$ is given by 
\begin{equation}
W_2(C|B)=W_2(\sigma |\eta ^{\prime \prime })=\prod_{i=1}^Nw_2(\sigma
_i|\eta _i^{\prime \prime }),  \label{a4}
\end{equation}
where 
\begin{equation}
\begin{array}{ll}
w_2(0|0)=1,        & \quad w_2(1|0)=0,  \\
w_2(2|0)=0,        & \quad w_2(3|0)=0.   \\
w_2(0|1)=0,        & \quad w_2(1|1)=1-p, \\
w_2(2|1)=\frac p2, & \quad w_2(3|1)=\frac p2, \label{a5}
\end{array}
\end{equation}

{\em Third stage}. In this state ($C\rightarrow A^{\prime }$), particles are
effectively created. Each occupied site remains occupied. Each vacant site
becomes occupied if the site at right (left) is occupied by a particle of
type $3$ ($2$).

A\ configuration of type $A^{\prime }$ is expressed again in terms of the
two state variables, $\eta _i=0$ or $1$, and the transition probability $%
W_2(A^{\prime }|C)$ of the transition from $C$ to $A^{\prime }$ is given by 
\begin{equation}
W_3(A^{\prime }|C)=W_3(\eta ^{\prime }|\sigma )=\prod_{i=1}^Nw_3(\eta
_i^{\prime }|\sigma _{i-1},\sigma _i,\sigma _{i+1}),  \label{a6}
\end{equation}
where the transition probability $w_3(1|\sigma _{i-1},\sigma _i,\sigma
_{i+1})$ to the state $\eta _i^{\prime }=1$ is given by 
\begin{equation}
w_3(1|\sigma _{i-1},\sigma _i,\sigma _{i+1})=1,  \label{a7}
\end{equation}
if $\sigma _i\neq 0,$ for any value of $\sigma _{i-1}$ and $\sigma _{i+1},$
and 
\begin{equation}
w_3(1|\sigma _{i-1},0,\sigma _{i+1})=1  \label{a8}
\end{equation}
if $\sigma _{i-1}=2$ or $\sigma _{i+1}=3.$ In other cases $w_3(1|\sigma
_{i-1},\sigma _i,\sigma _{i+1})$ vanishes. The transition probability $%
w_3(0|\sigma _{i-1},\sigma _i,\sigma _{i+1})$ to the state $\eta _i^{\prime
}=0$ is just given by 
\begin{equation}
w_3(0|\sigma _{i-1},\sigma _i,\sigma _{i+1})=1-w_3(1|\sigma _{i-1},\sigma
_i,\sigma _{i+1})  \label{a9}
\end{equation}

It ease to check that $W(A^{\prime }|A)=W(\eta ^{\prime }|\eta )$ cannot be
written as product of independent transition probability of each site as in
an ordinary cellular automaton. However, the transition probability 
$W_c(C^{\prime }|C)=W_c(\sigma ^{\prime }|\sigma )$ from state $C$ to 
state $C^{\prime }$ can. Indeed, from 
\begin{equation}
W_c(C^{\prime }|C)=\sum_{B^{\prime }}\sum_{A^{\prime }}W_2(C^{\prime
}|B^{\prime })W_1(B^{\prime }|A^{\prime })W_3(A^{\prime }|C)  \label{a10}
\end{equation}
that is from 
\begin{equation}
W_c(\sigma ^{\prime }|\sigma )=\sum_{\eta ^{\prime \prime }}\sum_{\eta
^{\prime }}W_2(\sigma ^{\prime }|\eta ^{\prime \prime })W_1(\eta ^{\prime
\prime }|\eta ^{\prime })W_3(\eta ^{\prime }|\sigma )  \label{a11}
\end{equation}
we get 
\begin{eqnarray}
W_c(\sigma ^{\prime }|\sigma )=\sum_{\eta ^{\prime \prime }}
\sum_{\eta^{\prime }}\prod_{i=1}^N&&
w_2(\sigma _i^{\prime }|\eta _i^{\prime \prime})
w_1(\eta _i^{\prime \prime }|\eta _i^{\prime })\times \nonumber \\
&&w_3(\eta _i^{\prime}|\sigma _{i-1},\sigma _i,\sigma _{i+1})  \label{a12}
\end{eqnarray}
which can be written in the form 
\begin{equation}
W_c(\sigma ^{\prime }|\sigma )=\prod_{i=1}^Nw(\sigma _i^{\prime }|\sigma
_{i-1},\sigma _i,\sigma _{i+1})  \label{a13}
\end{equation}
where 
\begin{eqnarray}
w(\sigma _i^{\prime }|\sigma _{i-1},\sigma _i,\sigma _{i+1})=
\sum_{\eta_i^{\prime \prime }}\sum_{\eta _i^{\prime }}&&
w_2(\sigma _i^{\prime }|\eta_i^{\prime \prime })
w_1(\eta _i^{\prime \prime }|\eta _i^{\prime })\times \nonumber \\
&&w_3(\eta_i^{\prime }|\sigma _{i-1},\sigma _i,\sigma _{i+1})  \label{a14}
\end{eqnarray}

The Grassberger and de la Torre process can then be viewed as sequence of
states $C,$ $C^{\prime },$ $C^{\prime \prime },$ $C^{\prime \prime \prime },
\cdots $, each one being given by a vector $\sigma =(\sigma _1,\sigma
_2,\cdots ,\sigma _N)$ where $\sigma _i=0,$ $1,$ $2,$ or $3$ according whether
site $i$ is either vacant, or occupied by a particle that does not generate
another particle (neutral particle), or occupied by a particle that
generates another one to the right (a rightist particle) or occupied by a
particle that generates another one to the left (a leftist particle).
Therefore, it is equivalent to an ordinary four-state cellular automaton
whose rules are defined by equations (\ref{a14}), (\ref{a3}), (\ref{a5}),
and (\ref{a7}-\ref{a8}). From these equations we may write down the
transition probability $w(\sigma _i^{\prime }|\sigma _{i-1},\sigma _i,\sigma
_{i+1})$ in the form 
\begin{equation}
\begin{array}{l}
w(0|\sigma _{i-1}^{\prime },\sigma _i^{\prime },\sigma _{i+1}^{\prime })=c,
\\ 
w(1|\sigma _{i-1}^{\prime },\sigma _i^{\prime },\sigma _{i+1}^{\prime })=a,
\\ 
w(2|\sigma _{i-1}^{\prime },\sigma _i^{\prime },\sigma _{i+1}^{\prime })=b/2,
\\ 
w(3|\sigma _{i-1}^{\prime },\sigma _i^{\prime },\sigma _{i+1}^{\prime })=b/2,
\end{array}
\label{a15}
\end{equation}
if $\sigma _i^{\prime }\neq 0,$ independently of the states taken by $\sigma
_{i-1}^{\prime },$ and $\sigma _{i+1}^{\prime }$. For the case where $\sigma
_i^{\prime }=0$, and either $\sigma _{i-1}^{\prime }=2$ or $\sigma
_{i+1}^{\prime }=3$, one has 
\begin{equation}
\begin{array}{l}
w(0|\sigma _{i-1}^{\prime },0,\sigma _{i+1}^{\prime })=c, \\ 
w(1|\sigma _{i-1}^{\prime },0,\sigma _{i+1}^{\prime })=a, \\ 
w(2|\sigma _{i-1}^{\prime },0,\sigma _{i+1}^{\prime })=b/2, \\ 
w(3|\sigma _{i-1}^{\prime },0,\sigma _{i+1}^{\prime })=b/2.
\end{array}
\label{a16}
\end{equation}
And finally, when $\sigma _i^{\prime }=0$, and $\sigma _{i-1}^{\prime }\neq
2 $ and $\sigma _{i+1}^{\prime }\neq 3$, 
\begin{equation}
\begin{array}{l}
w(0|\sigma _{i-1}^{\prime },0,\sigma _{i+1}^{\prime })=1, \\ 
w(1|\sigma _{i-1}^{\prime },0,\sigma _{i+1}^{\prime })=0, \\ 
w(2|\sigma _{i-1}^{\prime },0,\sigma _{i+1}^{\prime })=0, \\ 
w(3|\sigma _{i-1}^{\prime },0,\sigma _{i+1}^{\prime })=0.
\end{array}
\label{a17}
\end{equation}
The parameters $a$ and $b$ are related to $p$ by $a=(1-p)(1-c)$ and $%
b=p(1-c).$

The rule $w(0|0,0,0)=1$ implies that the state with all sites vacant is
indeed an absorbing state.

\section{acknowledgments}
J.~E.~S. would like to acknowledge financial support by FAPESP 
(Funda\c{c}\~ao de Amparo \`a Pesquisa do Estado de S\~ao Paulo).

%%%%%%%%%%%%%%%%%%%%%%%%%%%%%%%%%% REFERENCES %%%%%%%%%%%%%%%%%%%%%%%%%

\end{multicols}
\end{document}